# Optical absorption of divalent metal tungstates: Correlation between the band-gap energy and the cation ionic radius

R. Lacomba-Perales, J. Ruiz-Fuertes, D. Errandonea[*], D. Martínez-García, and A. Segura

Malta Consolider Team, Departamento de Física Aplicada-ICMUV, Universitat de València, Edificio de Investigación, C/Dr. Moliner 50, 46100 Burjassot (Valencia), Spain

**Abstract**

We have carried out optical-absorption and reflectance measurements at room temperature in single crystals of $AWO_4$ tungstates (A = Ba, Ca, Cd, Cu, Pb, Sr, and Zn). From the experimental results their band-gap energy has been determined to be 5.26 eV ($BaWO_4$), 5.08 eV ($SrWO_4$), 4.94 eV ($CaWO_4$), 4.15 eV ($CdWO_4$), 3.9 – 4.4 eV ($ZnWO_4$), 3.8 – 4.2 eV ($PbWO_4$), and 2.3 eV ($CuWO_4$). The results are discussed in terms of the electronic structure of the studied tungstates. It has been found that those compounds where only the $s$ electron states of the $A^{2+}$ cation hybridize with the O $2p$ and W $5d$ states (e.g $BaWO_4$) have larger band-gap energies than those where also $p$, $d$, and $f$ states of the $A^{2+}$ cation contribute to the top of the valence band and the bottom of the conduction band (e.g. $PbWO_4$). The results are of importance in view of the large discrepancies existent in prevoiusly published data.

**PACs Numbers:** 71.20.Nr, 78.40.Fy

---

[*] Corresponding author; email: daniel.errandonea@uv.es, Tel.: (34) 96 354 4475, Fax: (34) 354 3146

# I. Introduction

Metal tungstates (AWO$_4$) are semiconductors which usually crystallize in the tetragonal scheelite structure ($I4_1/a$) [1], for large A$^{2+}$ cations (A = Ba, Ca, Eu, Pb, Sr), or in the monoclinic wolframite structure ($P2/c$), for small A$^{2+}$ cations (A = Co, Cd, Fe, Mg, Ni, Zn) [2]. In scheelite W is coordinated by four O forming WO$_4$ tetrahedra while in wolframite W is surrounded by six O forming WO$_6$ octahedra. Other tungstates crystallize in structures related to scheelite, like monoclinic HgWO$_4$ ($C2/c$) [3] and cubic SnWO$_4$ ($P2_13$) [4], or to wolframite, like triclinic CuWO$_4$ ($P\bar{1}$) [5]. Nowadays, metal tungstates attract the attention of crystal-growth scientists, radiologists, and physicists due to their applications in the field of photonics and photoelectronics [6]. Their use in the detectors of the Large Hadron Collider at CERN [7], as laser-host materials [8], and in other optoelectronic devices like eye-safe Raman lasers [9] also created great interest on them. For all these applications an accurate knowledge of the band-gap energy (E$_g$) of tungstates is needed. However, despite the efforts made in the past, up to date no agreement concerning E$_g$ in metal tungstates has been obtained. The analysis of the literature data shows that the measured values of E$_g$ are widely dispersed (see Table I) [10 – 21]. In the particular cases of CaWO$_4$ [9] and ZnWO$_4$ [10], E$_g$ ranges from 4.4 to 6.8 eV and from 3.8 to 5.7 eV, respectively. It is, therefore, evident that E$_g$ in metal tungstates cannot be regarded as been accurately determined. In the present work we have measured E$_g$ for BaWO$_4$, SrWO$_4$, CaWO$_4$, CdWO$_4$, ZnWO$_4$, PbWO$_4$, and CuWO$_4$ by mean of reflectance and optical-absorption measurements. This approach has been previously probed to be very useful for obtaining accurately E$_g$ in many semiconductors [22].

**II. Experimental Details**

The samples used in the absorption experiments were thin platelets cleaved along natural cleavage directions of single crystals grown with the Czochralski method starting from commercial raw powders having 5N purity. More details on the crystal growth can be found elsewhere [7, 9, 23 – 25]. The thickness of the studied specimens ranged from 10 to 30 μm and their size was about 1000 μm x 1000 μm. In the reflectance measurements we used polished crystal plates 3 – 10 mm thick. The as grown crystals were colorless with the exception of the $CuWO_4$ crystal which had a dark-brown color. These crystals were characterized by x-ray diffraction. The obtained diffraction patterns corresponded to the structures reported in the literature for the seven tungstates. No indication of any extra phase was found in any of them. For the optical-absorption measurements in the UV-VIS-NIR we employed an optical set-up consistent of a deuterium lamp, fused silica lenses, reflecting optics objectives and an UV-VIS spectrometer, which allows for transmission measurements up to 5.5 eV [26]. The optical-absorption spectra were obtained from the transmittance spectra of the samples which were measured using the sample-in sample-out method [22]. The reflectance measurements were carried out at normal incidence.

**III. Results and discussion**

The absorption coefficient (α) of the seven studied compounds at room temperature is shown in Fig. 1. Given the thickness of the studied samples and the stray-light level of our spectroscopic system, the highest measurable value of the absorption coefficient is of the order of 2000 $cm^{-1}$, which is a typical value for the low-energy tails of direct-absorption edges. With the exception of $CuWO_4$, the absorption spectra of the compounds show similar features. They exhibit a steep absorption, characteristic of a direct band-gap, plus a low-energy absorption band which overlaps partially with the fundamental absorption. This low-energy absorption band has been previously observed in metal tungstates and seems to be related to the presence of defects or impurities [27]. Regarding, the steep absorption edge, as stated in the literature [12, 14, 21, 28], we found that it exhibits an exponential dependence on the photon energy following the Urbach's law [29]. This dependence is typical of the low-energy tails of direct-absorption edges with excitonic effects and has been attributed to the dissociation of excitons in the electric fields of polar phonons or impurities. The presence of the Urbach's tail is in agreement with the conclusion drawn from low-temperature measurements performed by Itoh *et al.* [12]. These authors as well as others also concluded that the lowest band-gap of metal tungstates is direct [12, 14, 19]. This result is also supported by recent electronic-structure calculations [21, 30] and contradicts the apparent indirect character assigned to the lowest band-gap of $CaWO_4$ and $SrWO_4$ by Arora *et al.* [10]. In the case of $CuWO_4$, the measured absorption spectra can be better explained assuming an indirect band-gap. This conclusion is confirmed by our reflectance measurements.

In order to determine $E_g$, we have analyzed the measured absorption spectra of the studied tungstates, but not $CuWO_4$, assuming that the band-gap is of direct type and that the absorption edge obeys the Urbach's rule $\alpha = A_0 e^{-(E_g - h\nu)/E_U}$ [29]. In this equation $E_U$ is the Urbach's energy, which is related to the steepness of the absorption tail, and $A_0 = k\sqrt{E_U}$ for a direct band-gap [18], being $k$ a characteristic parameter of each material. Fig. 2 illustrates the quality of the fits we obtained for our data using this model. As can be seen, the agreement of the fits with the experiments is quite good. For $CuWO_4$ we considered an indirect band-gap and $E_g$ was determined by plotting the square root of α and extrapolating a linear fit to zero (see Fig. 2). The results obtained by applying these analyses are summarized in Table I. In particular, we found that $E_g$ decreases following the sequence $BaWO_4 > SrWO_4 > CaWO_4 > CdWO_4 > ZnWO_4 > PbWO_4 > CuWO_4$. For $BaWO_4$, the determined $E_g$ agrees well with the values reported in the literature [19]. In the case of $CaWO_4$, the measured $E_g$ is 4.94 eV. This value is in agreement with that obtained by two-photon excitation techniques [11] suggesting that the band-gap energy of calcium tungstate is significantly lower than the previously accepted value (6.8 eV). For $CaWO_4$, Arora *et al.* obtained $E_g = 4.6$ eV from optical measurements similar to ours [10]. For $SrWO_4$ they obtained $E_g = 4.56$ eV, while we obtained $E_g = 5.08$ eV. We think that their underestimation of $E_g$ and their conclusion that $CaWO_4$ and $SrWO_4$ are indirect band-gap semiconductors were possibly caused by the fact that they used thick samples in their experiments. As a consequence of it, the defect-related low-energy tail could have been misinterpreted as a part of the fundamental absorption, leading to a different characterization of the band-gap and to an underestimation of its energy. In other compounds like $ZnWO_4$, $CdWO_4$, and $PbWO_4$ our results are also, like in $CaWO_4$, close to the lowest $E_g$ values found in the literature (see Table 1). In particular, for $PbWO_4$ our results agree with those reported by Itoh *et al.* [12, 14]. Regarding $CuWO_4$, we found $E_g = 2.3$ eV. This value is 1.3 eV smaller than the one previously reported by Arora *et al.* [17]. However, it is only 0.2 eV larger than the value obtained recently from thin films of $CuWO_4$ by Pandey *et al.* [16]. As the typical reduction of $E_g$ in thin films is about 0.2 eV [26] it seems that $E_g$ was previously overestimated in $CuWO_4$.

In the inset of Fig. 1 we show our reflectance measurements. For $CuWO_4$, we did not find any structure in the reflectance spectrum at the band-gap photon energy, which is consistent with the indirect character of its band-gap. For $PbWO_4$ and $ZnWO_4$, a clear maximum can be seen in the reflectance spectra. According with the position of these maxima we determined $E_g = 4.2$ and 4.4 eV, respectively. These values are slightly higher than the values obtained from the absorption measurements but still on the lower limit of the large dispersion of values reported for $E_g$ in the literature. In the cases of the alkaline-earth tungstates, the band-gap cannot be determined from the reflectance measurements since it is located very close to the high-photon energy limit of our spectrometer. As a consequence of it, there is no maximum present in the reflectance spectra (see inset of Fig. 1).

Density-functional theory [20], discrete variational [21], and *ab initio* [30] electronic-structure calculations indicate that in scheelite- and wolframite-structured tungstates the upper part of the valence band consists mainly of the $O^{2-}$ $2p$ states, and the conduction band is dominated by the $W^{6+}$ $5d$ states, in a similar way as it occurs in $WO_3$ [31]. The same conclusion was extracted for $SnWO_4$ [18]. The splitting caused by the crystal field in the $O^{2-}$ $2p$ states and the $W^{6+}$ $5d$ states produces the increase of $E_g$ with respect to $WO_3$ [20]. In addition, when the bivalent $A^{2+}$ cation of the tungstate belongs to group numbers 2 or 12 of the Periodic Table (i.e. the valence shell of A contains only $s$ electrons), the $s$ orbitals have some contribution to the valence and conduction bands. On the other hand, if A is a transition metal, an element of group number 14 of the Periodic Table, or a lanthanide, the $O^{2-}$ $2p$ states and the $W^{6+}$ $5d$ also hybridizes with $p$, $d$, or $f$ electrons of the $A^{2+}$ cation. Therefore the metal states contribution to the valence and conduction bands is more important. These conclusions have been confirmed by x-ray photoelectron spectroscopy measurements performed in $BaWO_4$, $CaWO_4$, $CdWO_4$, $ZnWO_4$, $PbWO_4$, and $CuWO_4$ [21, 32]. According with our results, the first group of bivalent metals leads to compounds with a larger $E_g$ than the second group of them (see Table I). We also observed that for both groups of tungstates a correlation can be established between the effective ionic (Shannon) radius [33] of the $A^{2+}$ metal and $E_g$. In Fig. 3 we plotted $E_g$ versus the Shannon radii of A. There it can be seen that for those elements whose valence shell contains only *s* electrons (Mg, Ca, Sr, Ba, Zn, Cd) there is a direct correlation between $E_g$ and the ionic radius; i.e., the bigger the $A^{2+}$ cation the larger the band-gap of the $AWO_4$ compound. A similar trend is followed by those compounds with bivalent cations with a different electronic configuration (Pb, Sn, Cu, Ni), but in such a case $E_g$ is systematically 1.4 eV smaller than in the former one. The smaller band-gap of the second group of compounds could be caused by the larger contribution of the metal states to the valence and conduction bands. Would be this hypothesis corrected, the pressure effects on the band-structure of the second group of compounds should be more important than in the first group. On the other hand, the correlation we established between $E_g$ and the ionic radius is coherent with the fact that electron states hybridization is expected to be affected by the size of the $A^{2+}$ cation [34].

As well, the correlation we found can be used to make a back-of-the-envelope estimation of the $E_g$ in other tungstates. We estimated $E_g$ = 4.55 eV ($HgWO_4$), 3.92 eV ($MgWO_4$), 3.70 eV ($EuWO_4$), 2.52 eV ($SnWO_4$), 2.52 eV ($NiWO_4$), 2.43 eV ($CoWO_4$), and 2.35 eV ($FeWO_4$). The predictions made for $MgWO_4$, $SnWO_4$, and $NiWO_4$ agree fairly well with previous experiments [15, 16, 18] supporting the correctness of the predictions made for the other four compounds, which however should de tested by future experiments and *ab initio* calculations. Finally, the lines drawn in Fig. 3 can be also used to predict $E_g$ in solid solutions of different tungstates (e.g. $Zn_xNi_{1-x}WO_4$) which as a first approximation can be assumed to vary linearly with the composition of the solid solutions [35].

## IV. Conclusions

In summary, the absorption and reflectance spectra of $BaWO_4$, $SrWO_4$, $CaWO_4$, $CdWO_4$, $ZnWO_4$, $PbWO_4$, and $CuWO_4$ have been accurately measured at room temperature. Our measurements suggest that all the studied compounds are direct band-gap semiconductors with the exception of CuWO$_4$ which is an indirect band-gap material. In addition, they allowed for a precise determination of E$_g$ in the seven studied tungstates and to solve preexistent discrepancies. We also found that in those compounds where the hybridization of the A metal states with W and O states is more important the band-gap is smaller than in those compounds where this hybridization is smaller. Based upon previous calculations, the reported results are explained in terms of the electronic structure of tungstates. Finally, a correlation between E$_g$ and the Shannon radii of the bivalent cation A was empirically found. This correlation is shown to be consistent with the present understanding of the electronic structure of tungstates and can be used to predict E$_g$ for unstudied tungstates like HgWO$_4$.

**Acknowledgements**: The authors thank P. Lecoq (CERN), P. Bohacek (Institute of Physics, Prague), and C.Y. Tu (Chinese Academy of Science) for providing AWO$_4$ samples. This study was supported by the Spanish MCYT (Grants No: MAT2007-65990-C03-01 and CSD2007-00045). D. E. and R. L.-P. acknowledge the financial support from the MCYT through the Ramón y Cajal and FPU programs.

**Table I:** $E_g$ and effective ionic radius [33] for different tungstates. (a) Measured values, (b) estimated values. The literature data were taken from Refs. [10 - 21].

| Compound | $E_g$ (eV) Literature | $E_g$ (eV) This work | Ionic radius of cation A (Å) |
|---|---|---|---|
| BaWO$_4$ | 4.8 – 5.2 | 5.26$^a$ | 1.42 |
| SrWO$_4$ | 4.56 | 5.08$^a$ | 1.26 |
| CaWO$_4$ | 4.4 – 6.8 | 4.94$^a$ | 1.12 |
| HgWO$_4$ |  | 4.55$^b$ | 1.02 |
| CdWO$_4$ | 4 – 5 | 4.15$^a$ | 0.95 |
| ZnWO$_4$ | 3.8 – 5.7 | 3.9 - 4.4$^a$ | 0.74 |
| MgWO$_4$ | 3.9 | 3.92$^b$ | 0.72 |
| PbWO$_4$ | 3.7 – 4.7 | 3.8 - 4.2$^a$ | 1.29 |
| EuWO$_4$ |  | 3.7$^b$ | 1.25 |
| CuWO$_4$ | 2.1 – 3.6 | 2.3$^a$ | 0.73 |
| SnWO$_4$ | 2.6 | 2.52$^b$ | 0.69 |
| NiWO$_4$ | 2.28 | 2.52$^b$ | 0.69 |
| CoWO$_4$ |  | 2.43$^b$ | 0.65 |
| FeWO$_4$ |  | 2.35$^b$ | 0.61 |

**Figure captions**

**Figure 1:** (color online) Absorption spectra of $BaWO_4$, $SrWO_4$, $CaWO_4$, $CdWO_4$, $ZnWO_4$, $PbWO_4$, and $CuWO_4$ single crystals. The inset shows the reflectance spectra.

**Figure 2:** Room temperature absorption spectra of $SrWO_4$ and $ZnWO_4$ showing the fits used to determine $E_g$. Dots: experiments. Lines: fits. The inset shows the $\alpha^{1/2}$ versus photon energy plot for $CuWO_4$ to illustrate the indirect character of its band-gap.

**Figure 3:** $E_g$ versus the effective ionic radius of the $A^{2+}$ cation in different tungstates. Circles: absorption measurements, squares: reflectance measurements, and triangles: average of value taken from Refs. [10 – 21]. Error bars represent the range of values reported for $E_g$ in the literature. Lines are just a guide to the eye.

**Figure 1**

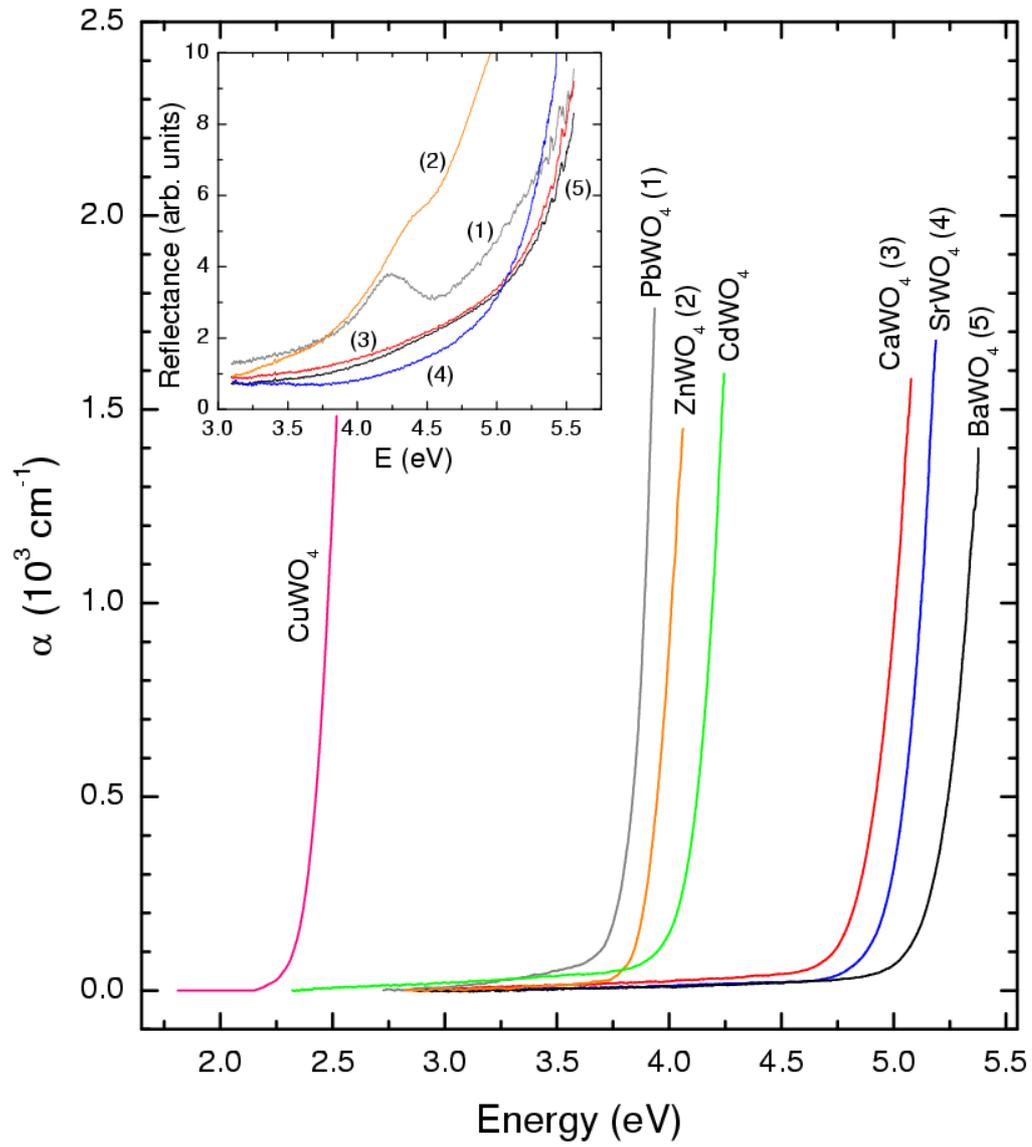

**Figure 2**

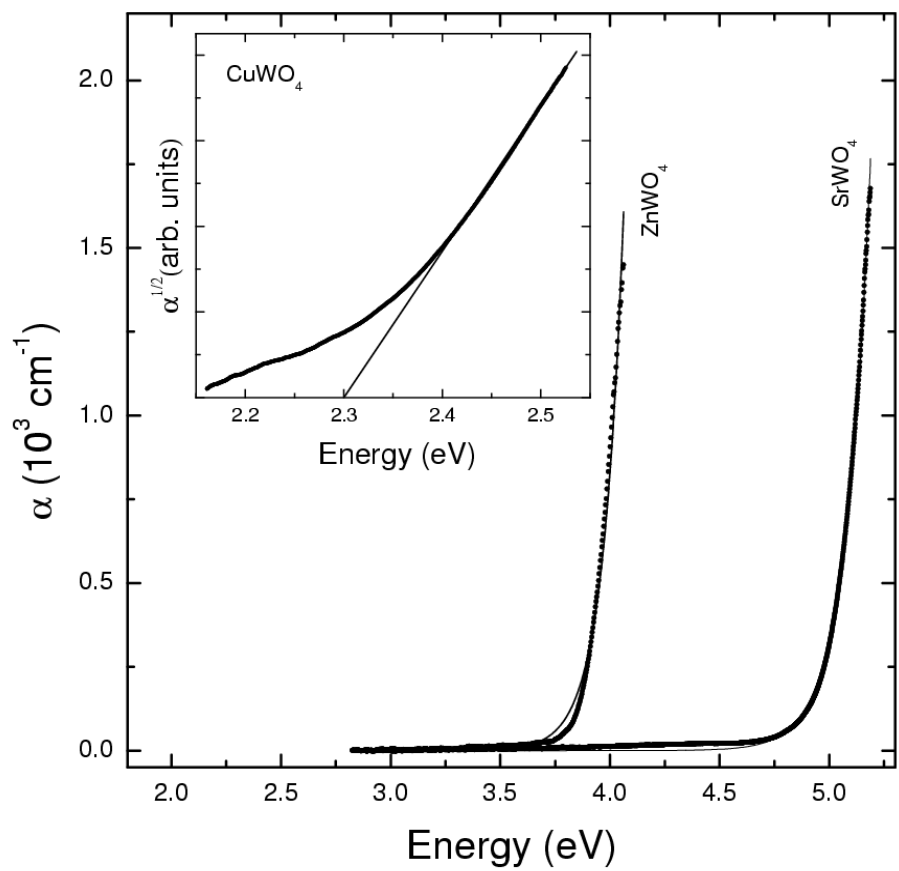

**Figure 3**

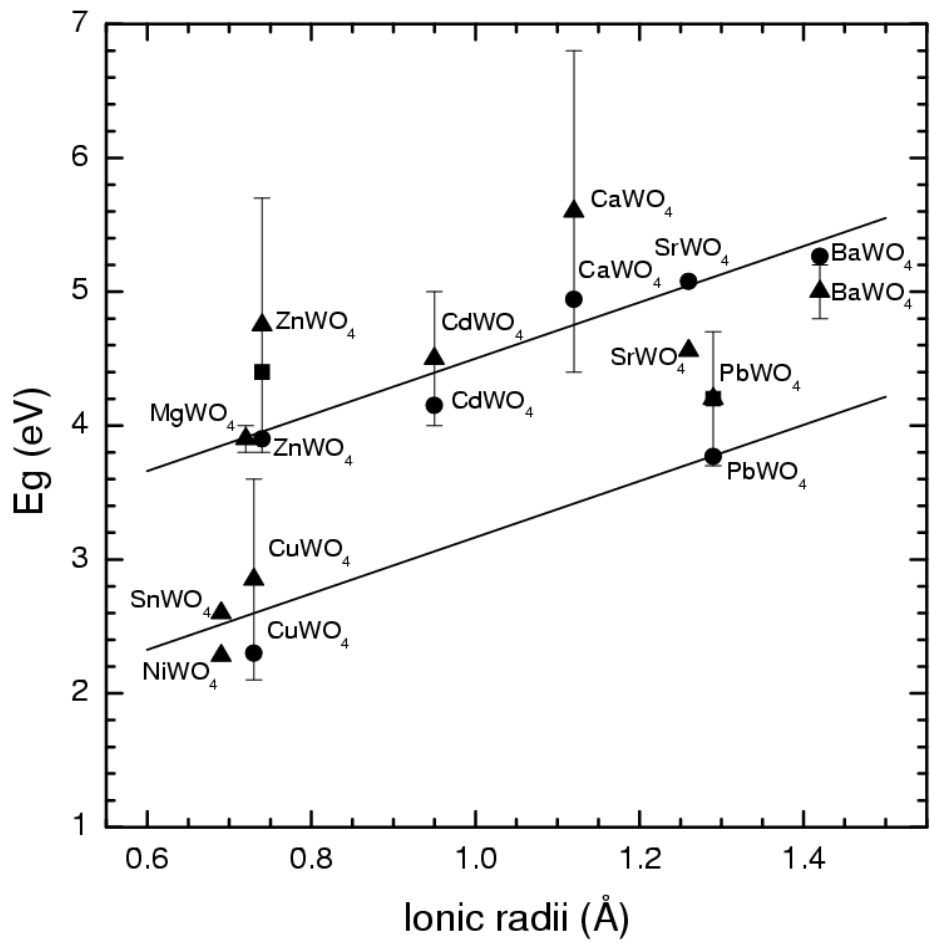